\documentclass[twoside]{article}
\usepackage{amsmath} 
\usepackage{amsfonts}
\usepackage{enumerate}  
\usepackage{dsfont}
\usepackage{qic-arxiv,epsfig}
\usepackage{verbatim}

\allowdisplaybreaks[1]

\usepackage{notation}

\newcommand\lref[1]{Lemma~\ref{lem:#1}}
\newcommand\tref[1]{Theorem~\ref{thm:#1}}
\newcommand\cref[1]{Corollary~\ref{cor:#1}}

\begin{document}
\setlength{\textheight}{8.0truein}    

\runninghead{A Complete Characterization of Pretty Good State Transfer on Paths}{C.M.~van Bommel}

\normalsize\textlineskip
\thispagestyle{empty}
\setcounter{page}{1}

\vspace*{0.88truein}

\alphfootnote

\fpage{1}

\centerline{\bf
A COMPLETE CHARACTERIZATION OF PRETTY}
\vspace*{0.035truein}
\centerline{\bf GOOD STATE TRANSFER ON PATHS}
\vspace*{0.37truein}
\centerline{\footnotesize
CHRISTOPHER M. VAN BOMMEL\footnote{Supported by a Canada Graduate Scholarship (Doctoral) from the Natural Sciences and Engineering Research Council of Canada.}}
\vspace*{0.015truein}
\centerline{\footnotesize\it {Department of Combinatorics and Optimization, University of Waterloo}}
\baselineskip=10pt
\centerline{\footnotesize\it Waterloo, Ontario, N2L 3G1, Canada}
\baselineskip=10pt
\centerline{\footnotesize\texttt{cvanbomm@uwaterloo.ca}}
\vspace*{0.225truein}
\centerline{\today}

\vspace*{0.21truein}

\abstracts{
We give a complete characterization of pretty good state transfer on paths between any pair of vertices with respect to the quantum walk model determined by the XY-Hamiltonian.  If $n$ is the length of the path, and the vertices are indexed by the positive integers from 1 to $n$, with adjacent vertices having consecutive indices, then the necessary and sufficient conditions for pretty good state transfer between vertices $a$ and $b$ are that (a) $a + b = n + 1$, (b) $n + 1$ has at most one odd non-trivial divisor, and (c) if $n = 2^t r - 1$, for $r$ odd and $r \neq 1$, then $a$ is a multiple of $2^{t - 1}$.
}{}{}

\vspace*{10pt}


\vspace*{1pt}\textlineskip 
 
\section{Introduction and Preliminaries}

Many quantum algorithms may be modelled as a quantum process occurring on a graph. In \cite{ChildsUniversalQComputation}, Childs shows that any quantum computation can be encoded as a quantum walk in some graph and thus quantum walks can be regarded as a quantum computation primitive. 

The protocol for quantum communication through unmeasured and unmodulated spin chains was presented by Bose~\cite{B03}, and led to the interpretation of quantum channels implemented by spin chains as wires for transmission of states.  We model such a spin chain of $n$ interacting qubits by the graph of a path of $n$ vertices, denoted $P_n$, with the vertices labelled from 1 to $n$ corresponding to qubits and the edges $\{i, i + 1\}$, $1 \le i < n$ corresponding to their interactions.  These interactions are defined by a time-independent Hamiltonian; that is, a symmetric matrix that acts on the Hilbert space of dimension $2^n$.  We are concerned here with the $XY$-Hamiltonian,  whose action on the 1-excitation subspaces is equivalent to the action of the 01-symmetric adjacency matrix of $P_n$ on $\C^n$, and whose process evolves according to the transition matrix $U(t) := \exp(itA)$.  To be more specific, we consider the Hamiltonian
\begin{equation} H = \sum_{a = 1}^{n - 1} \sigma^x_{a \ a + 1}\sigma^y_{a \ a + 1}, \end{equation}
where $\sigma^x_{a \ a + 1}$ and $\sigma^y_{a \ a + 1}$ are the operators that apply the Pauli matrices $\sigma^x$ and $\sigma^y$ at the qubits located at vertices $a$ and $a + 1$, and act as the identity at all other qubits. The sum is over all pairs of vertices  that are edges of the underlying graph. Consequently, if $|u\rangle$ stands for the system state in which the qubit at vertex $u$ is at $|1\rangle$ and all others are at $|0\rangle$, then
\begin{equation} H |a\rangle = \begin{cases}
| 2 \rangle, & a = 1 \\
| n - 1 \rangle, & a = n \\
| a - 1 \rangle + | a + 1 \rangle, & a \not\in \{1, n\}.
\end{cases} \end{equation}
Because of this, the action of $H$ on the set $\{|a\rangle : a \in V(G)\}$ is equivalent to the action of the adjacency matrix $A(P_n)$ on the canonical basis of $\C^n$. In other words, $H$ can be block diagonalized, and one of its blocks is equal to $A(P_n)$. The subspace spanned by $\{|a\rangle : a \in V(G)\}$ is the $1$-excitation subspace of the Hilbert space, and the quantum dynamics restricted to this subspace corresponds to the scenario where one qubit, say the one at $a$, is initialized at $|1\rangle$ and all others at $|0\rangle$. Due to Schrödinger's equation
\begin{equation} |\langle a |\exp(\ii t H) | b \rangle |^2 \end{equation}
indicates the probability that the state $|1\rangle$ is measured at $b$ after time $t$.

We are concerned solely with the 1-excitation subspace. Let $\ee_u$ denote the vector of the canonical basis of $\C^{n}$ that is $1$ at the entry corresponding to vertex $u$ at the ordering of the rows and columns of $A(G)$. From the remarks above, if the system is initialized with state $|1\rangle$ at vertex $a$ and all others at $|0\rangle$, it follows that
\begin{equation} |\ee_a^T \exp(\ii t A) \ee_b|^2 \end{equation}
indicates the probability that the state $|1\rangle$ is measured at $b$ after time $t$.  

One of the major goals of quantum communication on spin chains is to transfer a state with high fidelity.  At the maximum fidelity of 1, we say we have achieved \emph{perfect state transfer}.  Analogously, we say we have perfect state transfer between vertices $a$ and $b$ if there exists a time $\tau$ such that $|| \ee_a^T U(\tau) \ee_b|| = 1$.  The concept of perfect state transfer was first introduced by Christandl et al.~\cite{CDDEKL05}, who also showed perfect state transfer is only possible on spin chains of two or three qubits.  However, even without perfect state transfer, the fidelity may be quite high, and the notion of \emph{pretty good state transfer} was isolated by Godsil~\cite{G12} as a relaxation of perfect state transfer.  Formally, we say there is pretty good state transfer if, for any $\epsilon > 0$, there exists a time $\tau$ such that the fidelity is $1 - \epsilon$.  Analogously, we say we have pretty good state transfer between vertices $a$ and $b$ if, for any $\epsilon > 0$, there exists a time $\tau$ such that $|| U(\tau)_{a,b} || > 1 - \epsilon$, or equivalently, if for any $\epsilon > 0$, there exists a $\lambda \in \C$, $|\lambda| = 1$, such that $|| U(\tau) \ee_a - \lambda \ee_b || < \epsilon$; this condition is abbreviated to $U(\tau) \ee_a \approx \lambda \ee_b$ for convenience.  Godsil et al.~\cite{GKSS12} demonstrated the following result.

\vspace*{12pt}
\noindent
\begin{theorem} \cite{GKSS12} \label{thm:end}
Pretty good state transfer occurs between the end vertices of $P_n$ if and only if $n = p - 1, 2p - 1$, where $p$ is a prime, or $n = 2^m - 1$.   Moreover, when pretty good state transfer occurs between the end vertices of $P_n$, then it occurs between vertices $a$ and $n + 1 - a$ for all $a \neq (n + 1) /2$.
\end{theorem}
\vspace*{12pt}

Banchi et al.~\cite{BCGS16} showed that pretty good state transfer occurs between the $j$th and $(n - j +1)$-th vertices of spin chains with $XYZ$-Hamiltonian, whose action on the $1$-exication subspaces is equivalent to the action of the Laplacian adjacency matrix on $\C^n$, if the number of vertices is a power of 2, and that the condition is necessary for $j = 1$, but possibly not for other values of $j$.  Moreover, they present the open question of pretty good state transfer between inner vertices with $XY$-Hamiltonian.  Coutinho, Guo, and van Bommel~\cite{CGvB16} investigated this question and determined the following infinite family of paths that admit pretty good state transfer between inner vertices but not between the two end vertices.

\vspace*{12pt}
\noindent
\begin{theorem}\label{thm:main}\cite{CGvB16} Given any odd prime $p$ and positive integer $t$, there is pretty good state transfer in $P_{2^t p -1}$ between vertices  $a$ and $2^t p -a$, whenever $a$ is a multiple of $2^{t-1}$.
\end{theorem}
\vspace*{12pt}

In this paper, we present necessary and sufficient conditions for pretty good state transfer between any two vertices of a path, demonstrating that the above result is the only family of paths that admit pretty good state transfer between inner vertices but not between the two end vertices.  In the next section, we present definitions and preliminary results that will be required to prove this characterization.

\section{Preliminaries}

If $M$ is a symmetric matrix with $d$ distinct eigenvalues $\theta_1 > \theta_2 > \cdots > \theta_d$, then the spectral decomposition of $M$ is
\begin{equation} M = \sum_{j = 1}^d \theta_j E_j, \end{equation}
where $E_r$ denotes the idempotent projection onto the eigenspace corresponding to $\theta_r$.  If $a \in V(G)$, then the \emph{eigenvalue support} of $a$ is the following subset of the eigenvalues:
\begin{equation} \Theta_a = \{\theta_j : E_j \ee_a \neq 0\}. \end{equation}
We say that vertices $a$ and $b$ are \emph{strongly cospectral} if $E_r \ee_a = \pm E_r \ee_b$ for all $r$.  The following result is given by Banchi et al.~\cite{BCGS16}.

\vspace*{12pt}
\noindent
\begin{lemma} \cite{BCGS16} \label{lem:cospectral}
If pretty good state transfer occurs between $a$ and $b$, then they are strongly cospectral vertices.
\end{lemma}\vspace*{12pt}

The spectrum of the adjacency matrix of $P_n$ (see \cite{BH12} for example) is
\begin{equation} \theta_j = 2 \cos \frac{\pi j}{n + 1}, \quad 1 \le j \le n, \end{equation}
and the eigenvector corresponding to $\theta_j$ is given by
\begin{equation} (\beta_1, \beta_2, \ldots, \beta_k), \mbox{ where } \beta_k = \sin \frac{k \pi j}{n + 1}. \end{equation}
As stated by Coutinho, Guo, and van Bommel, the following lemma immediately follows.

\vspace*{12pt}
\noindent
\begin{lemma} \cite{CGvB16} \label{lem:sym}
Vertices $a$ and $b$ of $P_n$ are strongly cospectral if and only if $a + b = n + 1$.
\end{lemma}
\vspace*{12pt}

Moreover, we observe that when $a + b = n + 1$, $E_r \ee_a = E_r \ee_b$ when $r$ is odd, and $E_r \ee_a = - E_r \ee_b$ when $r$ is even.  

We will derive our next result, which gives a sufficient condition to show pretty good state transfer does not occur between a given pair of vertices of $P_n$, from Kronecker's Theorem, stated below.

\vspace*{12pt}
\noindent
\begin{theorem}[Kronecker, see \cite{LZ82}]
Let $\theta_0, \ldots, \theta_d$ and $\zeta_0, \ldots, \zeta_d$ be arbitrary real numbers.  For an arbitrarily small $\epsilon$, the system of inequalities
\begin{equation} | \theta_r y - \zeta_r | < \epsilon \pmod{2 \pi}, \quad (r = 0, \ldots, d), \end{equation}
admits a solution for $y$ if and only if, for integers $l_0, \ldots, l_d$, if
\begin{equation} \ell_0 \theta_0 + \cdots + \ell_d \theta_d = 0, \end{equation}
then
\begin{equation} \ell_0 \zeta_0 + \cdots + \ell_d \zeta_d \equiv 0 \pmod{2 \pi}. \end{equation}
\end{theorem}

\vspace*{12pt}
\noindent
\begin{lemma} \label{lem:nopgst}
Let $a$ and $b$ be vertices of $P_n$ such that $a + b = n + 1$.  If there is a set of integers $\{\ell_j : \theta_j \in \Theta_a,\ j \mbox{ odd}\}$ such that
\begin{equation} \sum_{\substack{\theta_j \in \Theta_a \\ j \mbox{ odd}}} \ell_j \theta_j = 0 \quad \mbox{and} \quad \sum_{\substack{\theta_j \in \Theta_a \\ j \mbox{ odd}}} \ell_j \mbox{ is odd} \end{equation}
and there is a set of integers $\{\ell_j : \theta_j \in \Theta_a,\ j \mbox{ even}\}$ such that
\begin{equation} \sum_{\substack{\theta_j \in \Theta_a \\ j \mbox{ even}}} \ell_j \theta_j = 0 \quad \mbox{and} \quad \sum_{\substack{\theta_j \in \Theta_a \\ j \mbox{ even}}} \ell_j \mbox{ is odd} \end{equation}
then pretty good state transfer does not occur between vertices $a$ and $b$.
\end{lemma}

\vspace*{12pt}
\noindent
{\bf Proof:} 
Suppose pretty good state transfer occurs between vertices $a$ and $b$.  We see that the condition $U(\tau) \ee_a \approx \lambda \ee_b$ is equivalent to $ e^{i \theta_r \tau} E_r \ee_a \approx \lambda E_r \ee_b$ for all $r$.  Writing $\lambda = e^{i \delta}$, this condition is equivalent to $\theta_r \tau \approx \delta + q_r \pi$ for all $r$ such that $\theta_r \in \Theta_a$, where $q_r$ is even when $r$ is odd and $q_r$ is odd when $r$ is even.  So, we wish to solve the system of inequalities
\begin{equation} |\theta_r \tau - (\delta + \sigma_r \pi) | < \epsilon \pmod{2 \pi}, \quad (r : \theta_r \in \Theta_a), \end{equation}
where $\sigma_r = 0$ when $r$ is odd and $\sigma_r = 1$ when $r$ is even. Hence, by Kronecker's Theorem, if
\begin{equation} \sum_{\theta_r \in \Theta_a} \ell_r \theta_r = 0, \end{equation}
then
\begin{equation} \sum_{\theta_r \in \Theta_a} \ell_r (\delta + \sigma_r \pi) \equiv 0 \pmod{2 \pi}. \end{equation}
Now, since there is a set of integers $\{\ell_j : \theta_j \in \Theta_a,\ j \mbox{ odd}\}$ such that
\begin{equation} \sum_{\substack{\theta_j \in \Theta_a \\ j \mbox{ odd}}} \ell_j \theta_j = 0 \quad \mbox{and} \quad \sum_{\substack{\theta_j \in \Theta_a \\ j \mbox{ odd}}} \ell_j \mbox{ is odd} \end{equation}
then we must have
\begin{equation} \delta \equiv 0 \pmod{2 \pi}.  \end{equation}
However, since there is a set of integers $\{\ell_j : \theta_j \in \Theta_a,\ j \mbox{ even}\}$ such that
\begin{equation} \sum_{\substack{\theta_j \in \Theta_a \\ j \mbox{ even}}} \ell_j \theta_j = 0 \quad \mbox{and} \quad \sum_{\substack{\theta_j \in \Theta_a \\ j \mbox{ even}}} \ell_j \mbox{ is odd} \end{equation}
then we must also have
\begin{equation} \delta \equiv \pi \pmod{2 \pi}, \end{equation}
which is the contradiction completing the proof \square\,.

\section{Necessary and Sufficient Conditions}

We first state the following results about sums of cosines which we will use in the proof of the main theorem.  Their proofs are included for completeness.  The first result uses the fact that $2 \cos x = e^{i x} + e^{-i x}$, and then we sum the resulting geometric series.

\vspace*{12pt}
\noindent
\begin{lemma}
Let $q$ be an odd integer.  Then
\begin{equation} 2 \sum_{k = 1}^{\frac{q - 1}{2}} (-1)^k \cos \left( \frac{k \pi}{q} \right) + 1 = 0. \end{equation}
\end{lemma}

\vspace*{12pt}
\noindent
{\bf Proof:} ~
\begin{align*}
& 2 \sum_{k = 1}^{\frac{q - 1}{2}} (-1)^k \cos \left( \frac{k \pi}{q} \right) + 1 \\
&= \sum_{k = 1}^{\frac{q - 1}{2}} (-1)^k (e^{\frac{i k \pi}{q}} + e^{\frac{- i k \pi}{q}}) + 1 \\
&= \sum_{k = 1}^{\frac{q - 1}{2}} (-e^{\frac{i \pi}{q}})^k + \sum_{k = 1}^{\frac{q - 1}{2}} (-e^{-\frac{i \pi}{q}})^k + 1 \\
&= \frac{-e^{\frac{i \pi}{q}} + (-e^{\frac{i \pi}{q}})^{\frac{q + 1}{2}}}{1 + e^{\frac{i \pi}{q}}} + \frac{(-e^{\frac{i \pi}{q}})^{-1} + (-e^{\frac{i \pi}{q}})^{-\frac{q + 1}{2}}}{1 + e^{-\frac{i \pi}{q}}} + 1 \\
&= \frac{-e^{\frac{i \pi}{q}} + (-e^{\frac{i \pi}{q}})^{\frac{q + 1}{2}}}{1 + e^{\frac{i \pi}{q}}} - \frac{1+ (-e^{\frac{i \pi}{q}})^{-\frac{q - 1}{2}}}{1 + e^{\frac{i \pi}{q}}} + \frac{1 + e^{\frac{i \pi}{q}}} {1 + e^{\frac{i \pi}{q}}}  \\
&= \frac{(-e^{\frac{i \pi}{q}})^{\frac{q + 1}{2}} - (-e^{\frac{i \pi}{q}})^{-\frac{q - 1}{2}}}{1 + e^{\frac{i \pi}{q}}} \\
&= 0 \square
\end{align*}

The second result makes use of the following sum-product identity:
\begin{equation} \cos \alpha + \cos \beta = 2 \cos \frac{\alpha + \beta}{2} \cos \frac{\alpha - \beta}{2} . \end{equation}
The final step is an application of the previous lemma.

\vspace*{12pt}
\noindent
\begin{lemma} \label{lem:acos}
Let $n = km$, where $m$ is an odd integer, and $0 \le a < k$ be an integer.  Then
\begin{equation} \sum_{j = 0}^{m - 1} (-1)^j \cos \left( \frac{(a + jk) \pi}{n} \right) = 0. \end{equation}
\end{lemma}

\vspace*{12pt}
\noindent
{\bf Proof:} 
\begin{align*}
& \sum_{j = 0}^{m - 1} (-1)^j \cos \left( \frac{(a + jk) \pi}{n} \right) \\
&= (-1)^{\frac{m - 1}{2}} \left[ \cos \left( \frac{\left(a + \frac{m - 1}{2} k \right) \pi}{n} \right) \right. \\ &\quad \left.+ \sum_{j = 1}^{\frac{m - 1}{2}} (-1)^j \left(\cos \left( \frac{\left(a + (\frac{m - 1}{2} - j) k \right) \pi}{n} \right) + \cos \left( \frac{\left(a + (\frac{m - 1}{2} + j) k \right) \pi}{n} \right) \right) \right] \\
&= (-1)^{\frac{m - 1}{2}} \left[ \cos \left( \frac{\left(a + \frac{m - 1}{2} k \right) \pi}{n} \right) \right. \\ &\quad \left.+ \sum_{j = 1}^{\frac{m - 1}{2}} (-1)^j 2 \cos \left( \frac{\left(a + \frac{m - 1}{2} k \right) \pi}{n} \right) \cos \left( \frac{j k \pi}{n} \right) \right] \\
&= (-1)^{\frac{m - 1}{2}} \cos \left( \frac{\left(a + \frac{m - 1}{2} k \right) \pi}{n} \right) \left( 1 + 2 \sum_{j = 1}^{\frac{m - 1}{2}} (-1)^j \cos \left( \frac{j \pi}{m} \right) \right) \\
&= 0 \square
\end{align*}

We will now state and prove the main theorem.

\vspace*{12pt}
\noindent
\begin{theorem}
There is pretty good state transfer on $P_n$ between vertices $a$ and $b$ if and only if $a + b = n + 1$ and either:
\begin{enumerate}
\item $n = 2^t - 1$, where $t$ is a positive integer; or,
\item $n = 2^t p - 1$, where $t$ is a nonnegative integer and $p$ is an odd prime, and $a$ is a multiple of $2^{t - 1}$.
\end{enumerate}
\end{theorem}

\vspace*{12pt}
\noindent
{\bf Proof:} 
The sufficiency of the conditions is given by \tref{end} and \tref{main}.  It remains to show that the conditions are necessary.  The necessity of the first condition follows from \lref{cospectral} and \lref{sym}.  Henceforth, we need only consider the possibility of pretty good state transfer between vertices $a$ and $n + 1 - a$.

Suppose first that there is pretty good state transfer on $P_n$ between vertices $a$ and $n + 1 - a$ when $n = 2^t r - 1$, where $t$ is a positive integer and $r$ is an odd composite number.  Let $p$ be a prime factor of $r$.  If $p \mid \frac{2^t r}{\gcd(a, 2^t r)}$, then if $\theta_k \notin \Theta_a$, then $k$ is a multiple of $p$.  But then, for $c \in \{1, 2\}$, we have
\begin{equation} \sum_{i = 0}^{r/p - 1} (-1)^i \theta_{c + i 2^t p} = \sum_{i = 0}^{r/p - 1} (-1)^i \cos \left( \frac{(c + i 2^t p) \pi}{n + 1} \right) = 0 \end{equation}
by \lref{acos}.  Hence, it follows from \lref{nopgst} that $P_n$ does not have pretty good state transfer between $a$ and $n + 1 - a$, a contradiction.

Hence, we can now assume that $r \mid a$.  Since it follows from condition (a) that $a \neq 2^{t-1} r$, we have that $t \ge 2$ and $4 \mid \frac{2^t r}{\gcd(a, 2^t r)}$, so if $\theta_k \notin \Theta_a$, then $k$ is a multiple of $4$.  But then, for $c \in \{1, 2\}$, we have
\begin{equation} \sum_{i = 0}^{r - 1} (-1)^i \theta_{c + i 2^t} = \sum_{i = 0}^{r - 1} (-1)^i \cos \left( \frac{(c + i 2^t) \pi}{n + 1} \right) = 0 \end{equation}
by \lref{acos}.  Hence, it follows from \lref{nopgst} that $P_n$ does not have pretty good state transfer between $a$ and $n + 1 - a$, a contradiction.  

Now suppose that there is pretty good state transfer on $P_n$ between vertices $a$ and $n + 1 - a$ when $n = r - 1$, where $r$ is an odd composite number.  Let $p$ be a prime factor of $r$.  If $p \mid \frac{2^t r}{\gcd(a, 2^t r)}$, then if $\theta_k \notin \Theta_a$, then $k$ is a multiple of $p$.  But then, for $c \in \{1, 2\}$, we have
\begin{align}
\notag &\sum_{i = 0}^{\frac{1}{2} (r/p - 1)} \theta_{c + i 2 p} + \sum_{i = 0}^{\frac{1}{2} (r/p - 1) - 1} \theta_{n + 1 - (c + p + i 2 p)} \\
&= \sum_{i = 0}^{r/p - 1} (-1)^i \theta_{c + i p} = \sum_{i = 0}^{r/p - 1} (-1)^i \cos \left( \frac{(c + i p) \pi}{n + 1} \right) = 0
\end{align}
by \lref{acos}.  Hence, it follows from \lref{nopgst} that $P_n$ does not have pretty good state transfer between $a$ and $n + 1 - a$, a contradiction.  Hence, we have shown the necessity of the conditions on $n$.

It remains to show the necessity of the conditions on $a$ when $n = 2^t p - 1$, where $t$ is a positive integer and $p$ is an odd prime.  Suppose $a$ is not a multiple of $2^{t - 1}$.  Then again we have that if $\theta_k \notin \Theta_a$, then $k$ is a multiple of $4$.  So as above, it follows from \lref{nopgst} that $P_n$ does not have pretty good state transfer between $a$ and $n + 1 - a$, which is the contradiction completing the proof \square\,.

\nonumsection{Acknowledgements}
\noindent
The author thanks Chris Godsil for his guidance and support and Gabriel Coutinho for his introduction to this problem.

\nonumsection{References}

\end{document}